# Broadband Cross-Circular Polarization Carpet Cloaking based on a Phase Change Material Metasurface in the Mid-infrared Region

Bo Fang[1], Dantian Feng[2,3], Peng Chen[1], Lijiang Shi[4], Jinhui Cai[1], Jianmin Li[5], Chenxia Li[2,&], Zhi Hong[3], Xufeng Jing[2,3,*]

[1]College of Metrology & Measurement Engineering, China Jiliang University, Hangzhou 310018, China
[2]Institute of Optoelectronic Technology, China Jiliang University, Hangzhou 310018, China
[3]Centre for THz Research, China Jiliang University, Hangzhou 310018, China
[4]Hangzhou Hangxin Qihui Technology Co., Ltd, Hangzhou, China
[5]University of Shanghai for Science and Technology, No.516 Jung Gong Road, Shanghai 200093, China
Corresponding authors: [4]lichenxia@cjlu.edu.cn, [*]jingxufeng@cjlu.edu.cn

*Abstract*: **In view of the fact that most invisibility devices focus on linear polarization cloaking and that the characteristics of mid-infrared cloaking are rarely studied, we propose a cross-circularly polarized invisibility carpet cloaking device in the mid-infrared band. Based on the Pancharatnam–Berry phase principle, the unit cells with the cross-circular polarization gradient phase were carefully designed and constructed into a metasurface. In order to achieve tunable cross-circular polarization carpet cloaks, a phase change material is introduced into the design of the unit structure. When the phase change material is in amorphous and crystalline states, the proposed metasurface unit cells can achieve high-efficiency cross-polarization conversion and reflection intensity can be tuned. According to the phase compensation principle of carpet cloaking, we construct a metasurface cloaking device with a phase gradient using the designed unit structure. From the near- and far-field distributions, the cross-circular polarization cloaking property is confirmed in the broadband wavelength range of 9.3–11.4 μm. The proposed cloaking device can effectively resist detection of cross-circular polarization.**
*Keywords*—**Metamaterial, Metasurface, Cloaking**

## I. INTRODUCTION

CLOAKING has always been a dream of humans, and the emergence of metamaterials has ushered in a new era of stealth technology [1-4]. Because metamaterials have a unique ability to control electromagnetic waves [5-20], scientists have strong interest in them [21-39] and adopt various principles and methods to design stealth devices. In 2006, Pendry's team and U. Leonhardt independently developed the concept and theory of transformation optics, making stealth devices a real possibility [40,41]. The theory of transformation optics proves that metamaterials with gradient parameters can control electromagnetic waves around obstacles, thus forming a cloaking space. However, it is difficult to design and fabricate a transformation optical stealth device because it requires a strict distribution of equivalent dielectric parameters. Also, a scattering cancellation invisibility cloak was proposed [42,43], which is robust to manufacturing tolerances. However, these cloaks are typically limited to sub wavelength objects. In addition, there are diverse invisibility cloaks, such as scattering reconstruction invisibility [44,45], near-zero refractive [46], reconfigurable acoustic [47], topology optimized [48], intelligent self-adaptive invisibility [49], and so on. Recently, researchers have proposed metasurface cloaks, which represent a major advance due to their ultrathin structure and ability to manipulate electromagnetic waves [50]. However, most of the aforementioned cloaking techniques focus on linearly polarized light and targeted visible, terahertz, and microwave wavelengths.

Circularly polarized antennae can receive arbitrarily polarized electromagnetic waves, and emitted electromagnetic waves can also be received by arbitrarily polarized antennae. Circularly polarized antennae are widely used in radar detection systems. In order to detect circularly polarized antennae, a circularly polarized cloaking device is proposed in our work. Polarization information is an important information target [51-53] and is of great significance for the analysis of polarization characteristics of a stealth target. A cross-polarized strong scatterer system can expand the polarization characteristics of electromagnetic waves to improve the signal-to-noise ratio and accurately detect and identify targets. In order address cross-polarization detection in a monitoring system, a cross-polarization cloaking device is proposed that uses Pancharatnam–Berry (PB) phase metasurfaces in the mid-infrared region.

Generally, once traditional metasurface cloaking devices are fabricated, the structure is unchangeable; its optical characteristics will not change, and the corresponding phase modulation function is also fixed. In order to flexibly control metasurfaces, researchers have introduced phase change materials in the design and fabrication of metasurfaces. Typical phase change materials include GeSbTe (GST) alloys [54], VO2 [55], graphene [56], and so on. The addition of phase change materials makes the optical properties of metasurface devices tunable. In order to achieve a compatible cross circular polarization stealth device, we introduce a GST phase change material into the metasurface cloaking design. The GST phase change material has significantly different optical properties between its crystalline and amorphous states. In this work, the classic metal-insulator-metal configuration with a GST layer is used to construct metasurfaces. Based on the principle of the PB phase [57], a switchable metasurface carpet cloak with amplitude regulation that can realize object-shielding functionality in the broadband wavelength range of 9.3–11.4 µm is proposed.

In order to realize the geometric phase characteristics of the cell structure, we need to design an anisotropic structure and apply it to the stealth device we constructed. To simplify our design, we propose a metal strip anisotropic structure to realize the geometric phase element. It should be noted that Li *et al.* proposed a multiple function of the meta-microstructure by combining a programmable metasurface with a radiation array, which can manipulate the scattering properties, digitally and in real-time, and exhibit different radiation modes simultaneously [58]. Also, they proposed a thin self-feeding Janus metasurface for manipulating incident waves and emitting radiation waves simultaneously [59]. By comparing the structure proposed by Li *et al.* and the structure designed by us, the structure designed by us has greater simplification and fewer structural parameters, which is conducive to the simplification of the design. It should be noted that the composite structure proposed by Li *et al.* is tunable and has many geometric parameters that can be used to achieve multifunctional characteristics.

## II. CLOAKING PRINCIPLE OF A METASURFACE

The generalized Snell's law is the basic principle to realize the cloaking effect in a carpet metasurface [60-64]. The generalized Snell's law is different from Snell's law, in which the results of refraction and reflection depend not only on different refractive indices in two media and the incident angle of incident wave, but also on the phase gradient function of a reflection surface. If a reflector with a specific phase gradient is designed, the phase and propagation direction of the reflected wave can be modulated. For example, the phase of the reflected wave of a convex object can be modulated to the phase of a reflected wave similar to that of the plane reflection in order to achieve the cloaking effect of the object underneath.

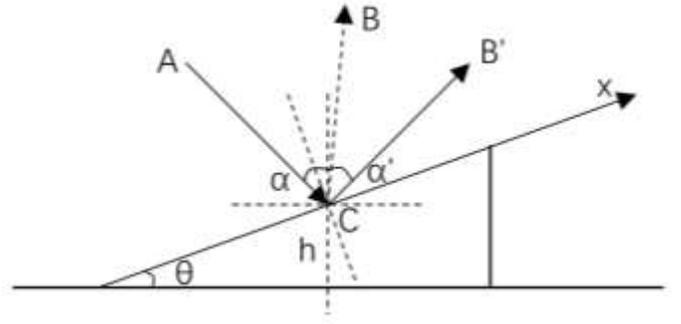

Fig.1. Schematic diagram of light propagation when incident light hits the left side of a carpet surface cloaking device

The generalized Snell's law for reflection can be expressed as [7]

$$sin\,\theta_r - sin\,\theta_i = \frac{\lambda_0}{2\pi n_i}\frac{d\phi}{dx}, \quad (1)$$

where $\lambda_0$ is the wavelength of incident light in the vacuum, $\theta_i$ is the incident angle, $\theta_r$ is the reflective angle, $n_i$ is the refractive index of the medium in incident space, and $\frac{d\phi}{dx}$ is the phase gradient of the interface. Anomalous reflection occurs when a beam of light is incident on a sloping slope covered with a carefully designed phase gradient metasurface, as shown in Fig.1. A schematic diagram of light propagation when incident light hits the left surface of a carpet surface cloaking device is shown in Fig.1. When the light emitted from point A is incident on point C on the inclined slope, the reflected light should be symmetric on the normal slope and exit from CB. However, due to the presence of the metasurface, the reflected light is modulated in the CB' direction, i.e., it is symmetrical to the horizontal normal. When $n_i = 1.0$, Eq.(1) can be deduced as

$$sin\,\theta_r - sin\,\theta_i = \frac{1}{k_0}\frac{d\phi}{dx}. \quad (2)$$

Based on the geometric relations in Fig.1,

$$\begin{cases} \theta_r = \alpha + \theta \\ \theta_i = \alpha - \theta \end{cases}, \quad (3)$$

where $\alpha$ is the angle formed by the incident ray and the vertical direction, and $\theta$ is the inclination angle of the inclined slope. We can further obtain

$$d\phi = 2k_0\,cos\,\alpha\,sin\,\theta\,dx. \quad (4)$$

With the geometric relationship $dx\,sin\,\theta = dh$, Eq.(4) can be further deduced with the integral operation as

$$\phi = 2k_0 h cos\,\alpha. \quad (5)$$

Half-wave loss will occur when light is reflected from an optically thin to an optically dense medium. In order to achieve the cloaking effect, the phase distribution to modulate on the metasurface slope needs to be achieved as

$$\delta = \pi - 2k_0 h cos\,\alpha. \quad (6)$$

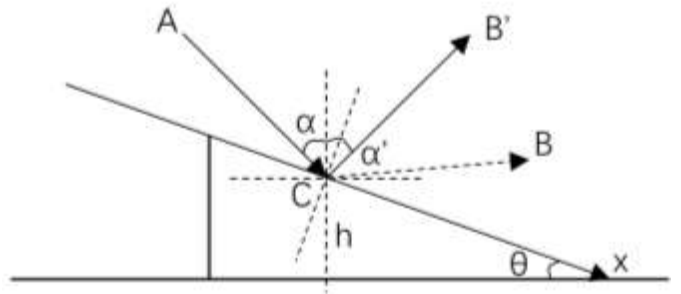

Fig.2. Schematic diagram of light propagation when incident light hits the right side of a carpet surface cloaking device

When a ray is incident on the right side of the phase gradient metasurface of the carpet cloaking device, the reflected ray will exhibit the abnormal reflection phenomenon shown in Fig.2. Similarly, the phase distribution to modulate on the metasurface slope can be achieved using Eq.(5). If a phase gradient metasurface that can provide specific phase compensation can be obtained, the reflected wave generated by incident light on the inclined slope can be modulated into a state similar to planar reflection so that a detector cannot detect its information, achieving the cloaking effect of hiding the object below the inclined plane.

## III. UNIT CELL OF A METASURFACE WITH PHASE CHANGE MATERIAL

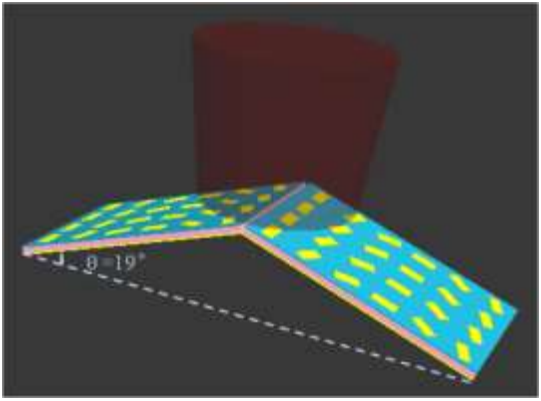

Fig.3. Schematic diagram of the metasurface carpet cloak

The key to designing a metasurface carpet stealth device that can realize dynamic control lies in the design of the unit structure. In order to dynamically control the stealth performance, a phase change material is introduced into the design of the unit cell. After obtaining the unit cells, we can arrange the units periodically to achieve a metasurface carpet cloak. A schematic of the metasurface carpet cloak is shown in Fig.3. The bottom inclination angle of the triangular carpet cloak is $\theta = 19°$, and circularly polarized light is perpendicularly incident on the carpet cloak. When the electromagnetic wave is incident on the metasurface carpet cloak, the appropriate phase shift is introduced by the metasurface unit cells, which change the phase of the local reflection, and the reflected wave front can be reconstructed as if the incident wave is incident on the interface of a plate. According to the principle of carpet cloaking, the local phase compensation introduced by the metasurface satisfies Eq.(6).

The metasurface unit cell that constitutes our carpet cloak consists of four layers of structure, and a schematic diagram of the unit cell is shown in Fig.4(a). The thickness of the bottom metal is $d_1$, and the middle two layers are the GST and MgF$_2$ thin film layers. As the phase change material layer, the GST layer is a key part to realize dynamic control of the stealth performance of the cloaking device. The MgF$_2$ layer serves as a refractive index matching layer between the high refractive index GST and the low refractive index air. This layer can improve the polarization conversion efficiency and prevent the GST film from oxidizing in the air. The thicknesses of the GST and MgF$_2$ layers are $d_2$ and $d_3$, respectively. The top layer is a metal patch layer with a thickness of $d_4$, and the length and width of the metal patch are $l$ and $w$, respectively. We analyze the controllable characteristics of the unit cell based on the GST phase change material and the period of the designed unit cell, which is $p = 4.2$ μm. Our tunable cloak is designed and optimized in the mid-infrared band. The dielectric constants of GST in amorphous and crystalline states are equal to 12 and 25, respectively, and the dielectric constant of MgF$_2$ is equal to 1.2 [63]. The detailed parameters of the optimized cell size are: $d_1 = 0.2$ μm, $d_2 = 0.6$ μm, $d_3 = 0.15$ μm, $d_4 = 0.1$ μm, $l = 2.6$ μm, and $w = 1.2$ μm. In the numerical simulation, the finite integral method of the commercial software CST microwave studio was applied to calculate optical characteristics of unit cells with co-polarized and cross-polarized reflectance. The unit cell boundary was used in the $x$ and $y$ directions. The open boundary was applied in the $z$ direction. The near- and far-field distributions of the designed cloaking device were numerically simulated using CST microwave studio. An open boundary was applied in the $x$, $y$, and $z$ directions. All the models described in our work were meshed using triangular elements with a maximum size of one-twentieth the wavelength.

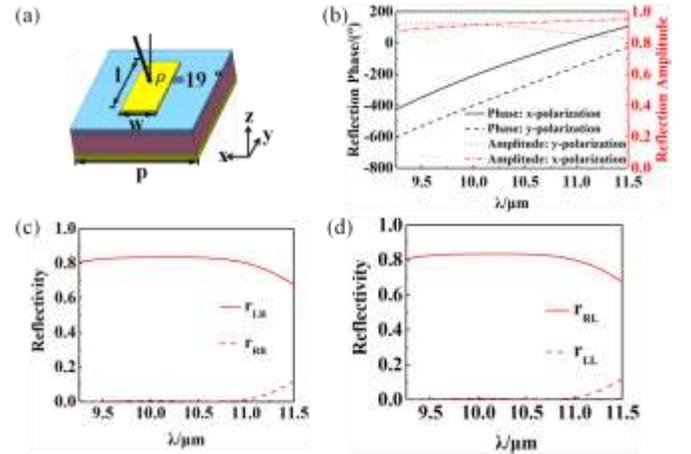

Fig.4. (a) Schematic diagram of the metasurface unit cell. (b) Amplitude and phase of the reflected wave under the incident condition of $x$- and $y$-polarized waves. Co-polarization and cross-polarization reflectivity under (c) LCP and (d) RCP incidence

Light is incident on the unit cell obliquely at an incident angle of $\rho = 19°$ (equal to the tilt angle of the bump). Fig. 4(b) shows the amplitude and phase of the reflected wave under the incidence conditions of $x$- and $y$-polarized waves. Here the GST material is in the amorphous state. The phase difference between two incident polarizations can be obtained as $|\Phi_x - \Phi_y| \approx 180°$ at a wavelength region of 9.3–11.4 μm. The reflection amplitudes of the unit cell under $x$- and $y$-polarized incident waves are both above 0.8. Figs. 4(c) and (d) show the reflectivity of co-polarization and cross-polarization under the incidence of left-handed circularly polarized (LCP) and right-handed circularly polarized (RCP) waves, respectively. In Figs. 4(c) and (d), $r_{RL}(r_{LL})$ and $r_{LR}(r_{RR})$ represent the cross-polarization (co-polarization) reflectivity when LCP and RCP waves are incident, respectively. It can be clearly seen from Figs. 4(c) and (d) that the proposed unit cell achieves high-efficiency cross-polarization conversion in the wavelength range of 9.3–11.4 μm. The cross-polarization reflectivity is above 0.8, and the co-polarization reflectivity is less than 0.1.

In order to achieve the stealth function of cross circular polarization, we need to obtain the cross-circular polarization reflection phase of the unit structure. According to the PB phase principle, different reflection phases with cross-polarization can be obtained by rotating a metal patch on the top layer of the unit. Taking six units as a cycle, the phase of the unit cell covers a range of $2\pi$, and the phase gradient of the unit cell is $\pi/3$. In view of the slight influence of the inclination angle of the stealth device ($\theta = 19°$), the required phase is obtained by scanning the rotation angle of the top metal patch. The parameters of the rotation angle, $\beta$, of the six units are shown in Table 1.

Table 1. Rotation angles and phase responses of the unit cells

| Unit cell | 1 | 2 | 3 | 4 | 5 | 6 |
|---|---|---|---|---|---|---|
| Rotation Angle | 0° | 31° | 62° | 90° | 120° | 150° |
| Phase | 0° | −60° | −120° | −180° | −240° | −300° |

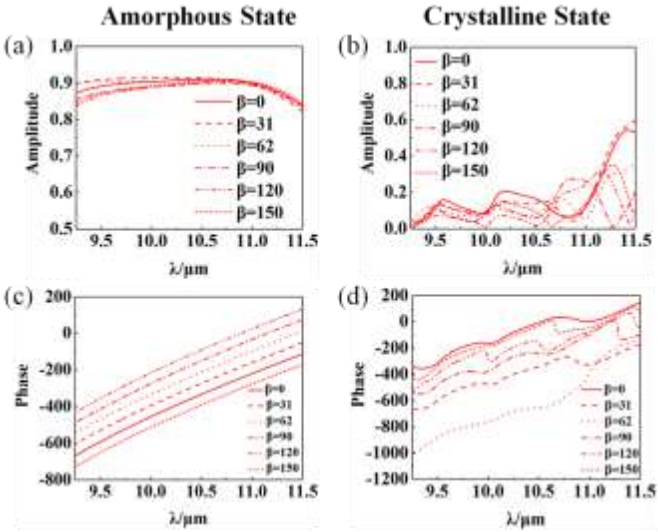

Fig.5. Amplitude and phase of the six units for cross-polarized reflection in the amorphous state and the crystalline state when the incident wave is LCP. The reflection amplitude for (a) amorphous and (b) crystalline states with different rotation angles. The reflection phase for (c) amorphous and (d) crystalline states

Figs. 5 (a) and (c) respectively show the reflection amplitude and phase of cross-polarization with the amorphous state of GST when the incident wave is LCP. Different rotation angles correspond to similar reflection amplitudes and different phase changes. Obviously, by controlling the local orientation of metallic elements between zero and $\pi$, phase variation covering the full $2\pi$ range can be realized while maintaining equal reflection amplitude. This characteristic of phase variation may also be called the geometric phase. The metasurface can be divided into geometric and resonant phase metasurfaces according to the phase mutation mechanism. In the early stage, we mainly studied the resonant phase metasurface, which moved the resonant frequency through the change of the structure, and then changed the phase of a certain frequency point, resulting in phase mutation. However, the resonant phase metasurface indicated some problems and defects since the phase mutation originates from the structural resonance, which leads to a limited working bandwidth of the resonant phase metasurface. A geometrical phase electromagnetic metasurface is a type of metasurface composed of the same artificial microstructure with different rotation angles. By simply changing the rotation angle of the microstructure, the phase mutation of reflected (transmitted) waves can be realized such that artificial control of the phase gradient or distribution is obtained. The geometrical phase is obtained from the rotational properties of the structure and is independent of wavelength [7, 8]. Therefore, a geometric phase metasurface has broadband characteristics.

When the GST is in the amorphous state, the cross-polarized reflection amplitudes of the six units in the wavelength range of 9.3–11.4 μm are all above 0.8, and the phase gradient between two adjacent unit cells is $\pi/3$, which meets the design requirement of the carpet cloak. Figs. 5 (b) and (d) are the reflection amplitude and phase of the cross-polarization of the six units when the GST is crystalline. The reflection amplitude of the unit structure is low, and the reflection phase is also variable with the rotation angle. We hope that by changing the state of the GST, the reflection intensity of the stealth device can be changed. Figs. 6 (a) and (c) are respectively the reflection amplitude and phase of the co-polarization of the six units when the GST is in the amorphous state. Figs. 6(b) and (d) are respectively the reflection amplitude and phase of the six units with different rotation angles when the GST is in the crystalline state. No matter if the GST material is in the amorphous or crystalline state, the amplitude and phase of the same polarization reflection does not change greatly with the change of rotation angle of the unit structure.

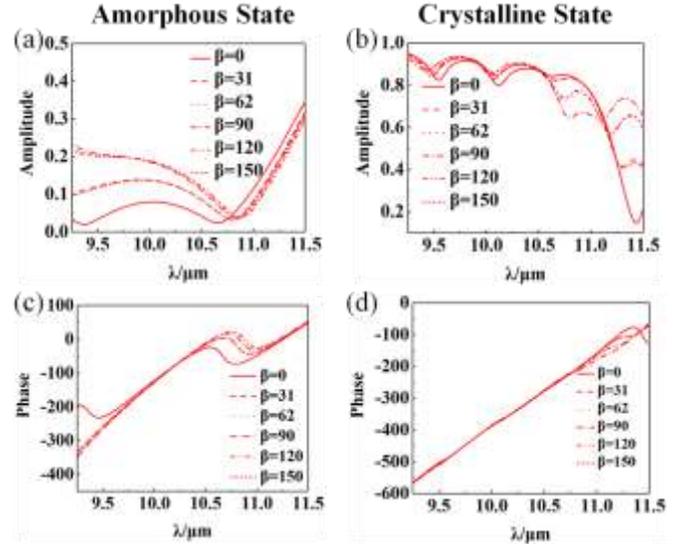

Fig.6. Amplitude and phase of the six units for co-polarized reflection in the amorphous and crystalline states, respectively, when the incident wave is LCP. (a) and (c) are for the amorphous state, and (b) and (d) are for the crystalline state

## IV. CROSS CIRCULAR POLARIZATION CARPET CLOAKING

Next, we use the designed unit structure to construct a carpet cloaking device, as shown in Fig. 3. We designed a stealth device with cross-circularly polarized stealth properties, and based on the amorphous and crystalline transition of the GST material, the designed stealth device has tunable reflection intensity. We can cover the metasurface unit cell on a triangular bump with an inclination angle of $\theta = 19°$ at the bottom. The height of the carpet cloak is 16.4 μm, and the length of both sides is 50.4 μm. The LCP is vertically incident on the designed stealth device. Fig. 7 illustrates the near-field distribution for a flat plate, triangular bulge, and metasurface carpet cloaks. Figs. 7(g), (h), and (i) show the near-field distributions when the GST is in the amorphous state, and Figs. 7(j), (k), and (l) show the near-field distributions when the GST is in the crystalline state. The near-field distributions of cross-polarized reflections at different incident wavelengths are also shown. For comparison, the near-field distribution of the flat plate is shown in Figs. 7 (a), (b), and (c), and the distribution of the exposed bumps, without covering the cloak, is shown in Figs. 7 (d), (e), and (f). When the incident wave hits the flat plate, the wave front of the reflected wave is completely a plane wave. When the incident wave hits the exposed bump that does not cover the carpet cloak, the wave front of the reflected wave is distorted. After the carpet cloak is covered on the bump, the wave front of the reflected wave returns to a plane wave when the GST is in the amorphous state. The near-field distribution is similar to that of an incident wave irradiated on the flat plate, which can successfully hide obstacles. When the GST is switched to the crystalline state, the reflected wave front also exhibits a plane wave characteristic. According to Fig. 5, the reflection intensity will vary with the state of the GST material.

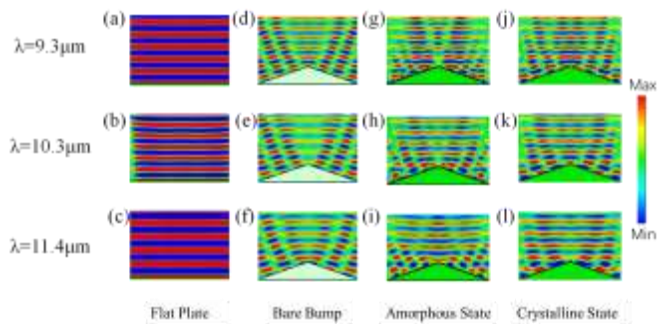

Fig.7. Near-field distribution of a (a), (b), and (c) flat plate; (d), (e), and (f), bare bump, (g), (h), and (i) amorphous state cloak; (j), (k), and (l) crystalline state cloak when λ = 9.3, 10.3, and 11.4 μm, respectively

In order to further demonstrate the cross-circularly polarized cloaking characteristics, we calculated the far-field scattering characteristics of the stealth device, as shown in Fig. 8. Figs. 8(g), (h), and (i) show the far-field scattering distribution with cross-circular polarization when the GST is in the amorphous state. Figs. 8(j), (k), and (l) show the far-field scattering distribution with cross-circular polarization when the GST is in the crystalline state. For comparison, the distribution of the flat plate is shown in Figs. 8 (a), (b), and (c), and the distribution of the exposed bumps, without covering the cloak, is shown in Figs. 8(d), (e), and (f). When the incident wave irradiates the exposed bump, most of the far-field energy is split into two beams. After covering the carpet cloak, the far-field scattering distribution returns to a distribution similar to the flat plate, and the energy is mainly concentrated in a single reflected beam with small side lobes. When the GST is in the crystalline state, the energy is still mainly concentrated in a single reflection beam, but the side lobes around 53° and 127° are more obvious and the stealth effect is slightly reduced. These results prove that the proposed carpet cloak has a good hiding effect when the GST is in crystalline or amorphous states.

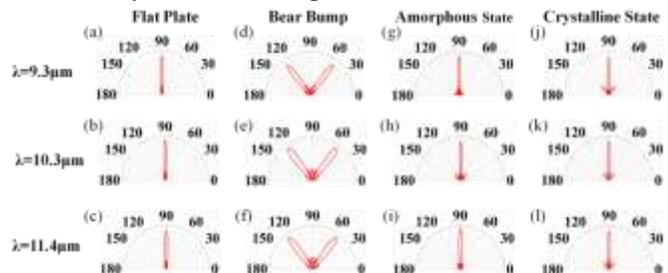

Fig.8. Far-field intensity distribution of (a), (b), and (c) flat plate; (d), (e), and (f) bare bump, (g), (h), and (i) amorphous state cloak; (j), (k), and (l) crystalline state cloak when λ = 9.3, 10.3, and 11.4 μm, respectively

The far-field radiation pattern in Fig. 9 more intuitively illustrates the broadband hiding effect of the designed metasurface carpet cloak. Figs. 9 (a), (b), (c), and (d) show the far-field radiation of the flat plate, bare bump, and carpet cloak when the GST is in the amorphous and crystalline states. It can be clearly seen that the designed metasurface carpet cloak can achieve a switchable invisibility effect in the broadband wavelength range of 9.3–11.4 μm.

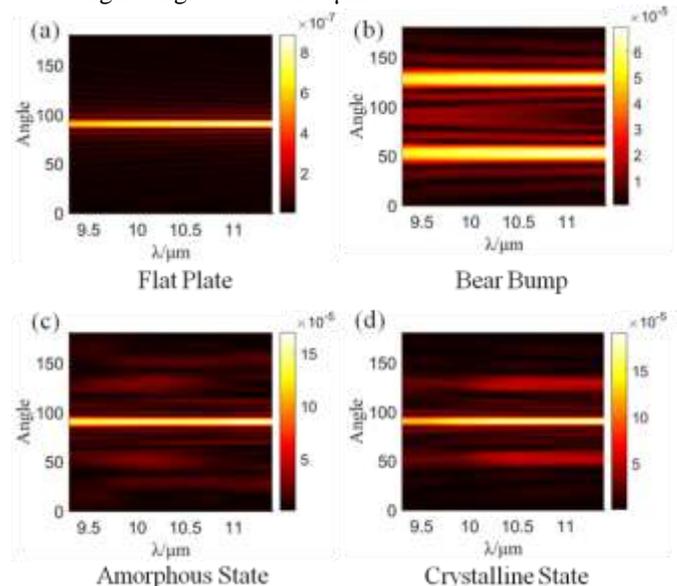

Fig.9. Far-field radiation pattern of a (a) flat plate, (b) bare bump, (c) amorphous state cloak, and (d) crystalline state cloak in the broadband range

Although our work is mainly focused on numerical simulations, it is feasible to fabricate and measure such metasurface cloaking devices. The designed cloaking metasurface can be fabricated as follows. First, the thin film metal, $MgF_2$, and GST can be deposited on a silicon dioxide substrate using magnetron sputtering and electron beam evaporation methods [65]. Photoresist can be spin coated onto the sample and baked at low temperatures. The metal rods can be patterned using laser direct-writing lithography to create a photoresist mask. Then, the photoresist pattern can be transferred to the top metal by ion beam etching. A fully cloaked structure with the fabricated metasurface can be covered with triangular bumps [61]. Morphologic analysis of the samples can be carried out using scanning electron microscopy and optical microscopy.

To measure cloaking performance, a $CO_2$ laser can be used as the light source. After passing through an adjustable attenuator and a linear polarizer, the optical beam can illuminate the fabricated cloaking device. A semi-transparent mirror can be applied in front of the device, and infrared charge-coupled devices can be utilized to obtain the reflection image. In order to measure the co-polarized and cross-polarized reflectance, Fourier transform infrared spectroscopy should be applied with a reflection accessory. Also, two linear polarizers should be utilized to generate broadband circularly polarized light [65].

To verify the accuracy of our numerical simulation results, we use COMSOL software for further calculations. Figure 10(a) shows the amplitude and phase of the reflected wave under the incidence conditions of $x$- and $y$-polarized waves. Figure 10(c) and (d) show the reflectivity of co-polarization and cross-polarization under the incidence of left-handed circularly polarized and right-handed circularly polarized waves, respectively. By comparing Fig. 10 and Fig. 4, we find that the results calculated by the two methods are consistent. These results confirm the accuracy of our design results.

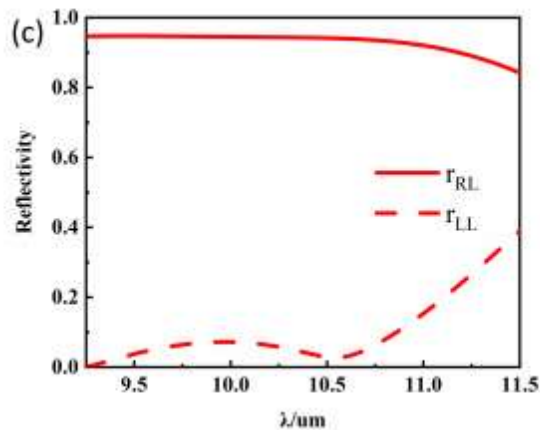

Fig.10. The results of numerical calculation by COMSOL software. (a) Amplitude and phase of the reflected wave under the incident condition of x- and y-polarized waves. Co-polarization and cross-polarization reflectivity under (b) LCP and (c) RCP incidence.

## V. Summary

In this paper, the phase change material GST is introduced into the design of a metasurface invisibility device. A broadband reflective adjustable metasurface carpet cloak for cross-polarization conversion of circularly polarized waves is proposed. According to the principles of phase compensation and the PB phase, the basic unit cell of a metasurface is designed. We use a six-unit structure for the period and arrange the structure onto a metasurface with a phase gradient to construct the metasurface stealth device. Based on the near- and far-field scattering distributions, the designed cross-circular polarization stealth device has a good stealth effect when the GST material is in crystalline and amorphous states.

**Acknowledgement**

This work was supported by National Key R&D Program of China (No. 2018YFF01013005); Natural Science Foundation of Zhejiang Province (No. LY22F050001, No. LZ21A040003, No. LY20F050007,); National Natural Science Foundation of China (No.52076200 and 62175224); New-shoot Talents Program of Zhejiang province (No.2021R409042 and No.2021R409012); The Fundamental Research Funds for the Provincial Universities of Zhejiang.

## VI. References

[1] X. Jing, C. Chu, C. Li, H. Gan, Y. He, X. Gui, and Z. Hong, Enhancement of bandwidth and angle response of metasurface cloaking through adding antireflective moth-eye-like microstructure, Opt.Express 27(15), 21767 (2019)

[2] L. Jiang, C. Chu, B. Fang, M. Zhang, H. Gan, C. Li, X. Jing, Z. Hong, Multi-wavelength carpet cloaking based on an ultrathin single layer metamaterial microstructure, Laser Phys.Lett. 17(6), 066202 (2020)

[3] J. Yang, H. Huang, X. Wu, B. Sun, X. Luo, Dual-Wavelength Carpet Cloak Using Ultrathin Metasurface, Adv.Opt.Mater. 6(14), 1800073 (2018)

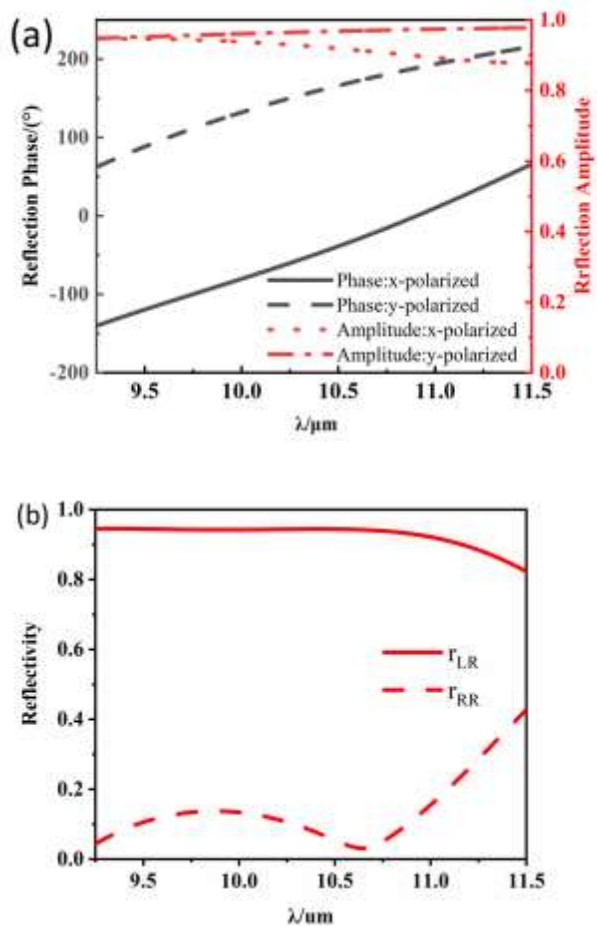

[4] J. Yang, S. Qu, H. Ma, J. Wang, S. Sui, Q. Zheng, H. Chen, Y. Pang, Ultra-broadband co-polarization anomalous reflection metasurface, Appl.Phys.A-Mater.Sci.Process. 123(8), 538 (2017)

[5] Jinxing Li, Yueyi Yuan, Qun Wu, Shah Nawaz Burokur, and Kuang Zhang, "Dual-band independent phase control based on high efficiency metasurface [Invited]," Chin. Opt. Lett. 19, 100501- (2021).

[6] S. Teng, Q. Zhang, H. Wang, L. Liu, and H. Lv, Conversion between polarization states based on metasurface, Photonics Res. 7(3), 246 (2019)

[7] M. Akram, G. Ding, K. Chen, Y. Feng, and W. Zhu, Ultra-thin single layer metasurfaces with ultra-wideband operation for both transmission and reflection, Adv.Mater 32, 1907308 (2020)

[8] J. Zhang, X. Wei, I. Rukhlenko, H. Chen, and W. Zhu, Electrically tunable metasurface with independent frequency and amplitude modulations, ACS Photonics 7(1), 265 (2020)

[9] Haoyu Wang, Zhiyu Zhang, Kun Zhao, Wen Liu, Pei Wang, and Yonghua Lu, "Independent phase manipulation of co- and cross-polarizations with all-dielectric metasurface," Chin. Opt. Lett. 19, 053601- (2021).

[10] B. Fang, Z. Cai, Y. Peng, C. Li, Z. Hong and X. Jing, Realization of ultrahigh refractive index in terahertz region by multiple layers coupled metal ring metamaterials, J.Electromagn.Waves Appl. 33(11), 1375 (2019)

[11] B. Fang, B. Li, Y. Peng, C. Li, Z. Hong, X. Jing, Polarization‐independent multiband metamaterials absorber by fundamental cavity mode of multilayer microstructure, Microw.Opt.Technol.Lett. 61(10), 2385 (2019)

[12] W. Wang, X. Jing, J. Zhao, Y. Li, Y. Tian, Improvement of accuracy of simple methods for design and analysis of a blazed phase grating microstructure, Opt.Appl. 47(2), 183 (2017)

[13] L. Jiang, B. Fang , Z Yan, et al. Improvement of unidirectional scattering characteristics based on multiple nanospheres array, Microw.Opt.Technol.Lett. 62(6), 2405 (2020)

[14] Yi Zhao, Qiuping Huang, Honglei Cai, Xiaoxia Lin, Hongchuan He, Hao Cheng, Tian Ma, and Yalin Lu, "Ultrafast control of slow light in THz electromagnetically induced transparency metasurfaces," Chin. Opt. Lett. 19, 073602- (2021).

[15] Xinhua Xie, Yunpei Deng, Steven L. Johnson. "Compact and robust supercontinuum generation and post-compression using multiple thin plates," High Power Laser Science and Engineering, 2021, 9(4): 04000e66.

[16] Ai Du, Yi Ma, Mingfang Liu, Zhihua Zhang, Guangwei Cao, Hongwei Li, Ling Wang, Peijian Si, Jun Shen, Bin Zhou. "Morphology analysis of tracks in the aerogels impacted by hypervelocity irregular particles," High Power Laser Science and Engineering, 2021, 9(2): 02000e14.

[17] Tina Ebert, René Heber, Torsten Abel, Johannes Bieker, Gabriel Schaumann, Markus Roth. "Targets with cone-shaped microstructures from various materials for enhanced high-intensity laser–matter interaction," High Power Laser Science and Engineering, 2021, 9(2): 02000e24.

[18] Hafiz Saad Khaliq, Inki Kim, Aima Zahid, Joohoon Kim, Taejun Lee, Trevon Badloe, Yeseul Kim, Muhammad Zubair, Kashif Riaz, Muhammad Qasim Mehmood, Junsuk Rho. "Giant chiro-optical responses in multipolar-resonances-based single-layer dielectric metasurfaces," Photonics Research, 2021, 9(9): 09001667.

[19] Matthew Parry, Andrea Mazzanti, Alexander Poddubny, Giuseppe Della Valle, Dragomir N. Neshev, Andrey A. Sukhorukov. "Enhanced generation of nondegenerate photon pairs in nonlinear metasurfaces," Advanced Photonics, 2021, 3(5): 055001.

[20] J. Zhang, H. Zhang, W. Yang, K. Chen, X. Wei, Y. Feng, R. Jin, and W. Zhu, Dynamic scattering steering with graphene-based coding meta-mirror, Adv.Opt.Mater. 8(1), 2000683 (2020)

[21] X. Bai, F. Kong, Y. Sun, F. Wang, J. Qian, X. Li, A. Cao, C. He, X. Liang, R. Jin, and W. Zhu, High-efficiency transmissive programable metasurface for multi-mode OAM generations, Adv.Opt.Mater. 8(1), 2000570 (2020)

[22] X. Jing, X. Gui, P. Zhou, and Z. Hong, Physical explanation of Fabry-Pérot cavity for broadband bilayer metamaterials polarization converter, J.Lightwave Technol. 36(12), 2322 (2018)

[23] R. Xia, X. Jing, X. Gui, and Y. Tian, Broadband terahertz half-wave plate based on anisotropic polarization conversion metamaterials, Opt.Mater.Express 7(3), 977, (2017)

[24] M. Akram, M. Mehmood, X. Bai, R. Jin, M. Premaratne, and W. Zhu, High efficiency ultra-thin transmissive metasurfaces, Adv.Opt.Mater. 7, 1801628 (2019)

[25] M. Akram, X. Bai, R. Jin, G. Vandenbosch, M. Premaratne, and W. Zhu, Photon spin Hall effect based ultra-thin transmissive metasurface for efficient generation of OAM waves, IEEE Trans.Antennas Propag. 67(7), 4650 (2019)

[26] J. Zhao, X. Jing, W. Wang, Y. Tian, D. Zhu, and G. Shi, Steady method to retrieve effective electromagnetic parameters of bianisotropic metamaterials at one incident direction in the terahertz region, Opt. Laser Technol. 95, 56 (2017)

[27] Y. Tian, X. Jing, H. Gan, X. Li and Z. Hong, Free control of far-field scattering angle of transmission terahertz wave using multilayer split-ring resonators' metasurfaces, Front.Phys. 15(6), 62502 (2020)

[28] C. Zhou, Z. Mou, R. Bao, Z. Li, and S. Teng, "Compound plasmonic vortex generation based on spiral nanoslits," Front. Phys. 16, 33503 (2021).

[29] G. Dai, "Designing nonlinear thermal devices and metamaterials under the Fourier law: A route to nonlinear thermotics," Front. Phys. 16, 53301 (2021).

[30] J. Li, R. Jin, J. Geng, X. Liang, K. Wang, M. Premaratne, and W. Zhu, Design of a broadband metasurface Luneburg lens for full-angle operation, IEEE Trans.Antennas Propag. 67 (4), 2442 (2019)

[31] X. Lu, X. Zeng, H. Lv, Y. Han, Z. Mou, C. Liu, S. Wang and S. Teng, Polarization controllable plasmonic focusing based on nanometer holes, Nanotechnology 31(13), 135201 (2020)

[32] H. Lv, X. Lu, Y. Han, Z. Mou, C. Zhou, S. Wang, and S. Teng, Metasurface cylindrical vector light generators based on nanometer holes, New J.Phys. 21(12), 123047 (2019)

[33] H. Lv, X. Lu, Y. Han, Z. Mou, and S. Teng, Multifocal metalens with a controllable intensity ratio, Opt.Lett. 44(10), 2518 (2019)

[34] H. Wang, L. Liu, C. Zhou, J. Xu, M. Zhang, S. Teng, and Y. Cai, Vortex beam generation with variable topological charge based on a spiral slit, Nanophotonics 8(2), 317 (2019)

[35] X. Jing, S. Jin, Y. Tian, P. Liang, Q. Dong, and L. Wang, Analysis of the sinusoidal nanopatterning grating structure, Opt.Laser Technol. 48, 160 (2013)

[36] X. Jing, Y. Xu, H. Gan, Y. He and Z. Hong, High Refractive Index Metamaterials by Using Higher Order Modes Resonances of Hollow Cylindrical Nanostructure in Visible Region, IEEE Access 7 144945 (2019)

[37] L. Jiang, B. Fang, Z. Yan, J. Fan, C. Qi, J. Liu, Y. He, C. Li, X. Jing, H. Gan, Z. Hong, Terahertz high and near-zero refractive

index metamaterials by double layer metal ring microstructure, Opt. Laser Technol. 123, 105949 (2020)

[38]  X. He, Tunable terahertz graphene metamaterials, Carbon 82, 229 (2015)

[39]  X. He, X. Zhong, F. Lin, and W. Shi, Investigation of graphene assisted tunable terahertz metamaterials absorber, Opt.Mater.Express 6(2), 331 (2016)

[40]  J. Pendry and D. Schurig and D. Smith, Controlling Electromagnetic Fields, Science 312(5781), 1780 (2006)

[41]  U. Leonhardt, Optical Conformal Mapping, Science 312(5781), 1777 (2006)

[42]  D. Deslandes and K. Wu, Accurate modeling, wave mechanisms, and design considerations of a substrate integrated waveguide, IEEE T.Microw.Theory. 54(6), 2516 (2006)

[43]  A. Rajput and K. Srivastava, Dual-band cloak using microstrip patch with embedded u-shaped slot, IEEE Antennas Wirel.Propag.Lett. 16, 2848 (2017)

[44]  Y. Yang, H. Wang, F. Yu et al., A metasurface carpet cloak for electromagnetic acoustic and water waves, Sci Rep 6, 20219 (2016)

[45]  J. Zhang, Z. Mei, W. Zhang et al., An ultrathin directional carpet cloak based on generalized Snell's law, Appl.Phys.Lett. 103(15), 151115 (2013)

[46]  S. Islam, M. IqbalFaruque, and M. Islam, A near zero refractive index metamaterial for electromagnetic invisibility cloaking operation, Materials 8(8), 4790 (2015)

[47]  S. Fan, S. Zhao, L. Cao et al., Reconfigurable curved metasurface for acoustic cloaking and illusion, Phys.Rev.B 101(2), 024104 (2020)

[48]  L. Lan, F. Sun, Y. Liu et al., Experimentally demonstrated a unidirectional electromagnetic cloak designed by topology optimization, Appl.Phys.Lett. 103(12), 121113 (2013)

[49]  M. Selvanayagam and G. Eleftheriades, Experimental demonstration of active electromagnetic cloaking, Phys.Rev.X 3(4), 041011 (2013)

[50]  C. Qian, B. Zheng, Y. Shen et al., Deep-learning-enabled self-adaptive microwave cloak without human intervention, Nat.Photonics 14(6), 383 (2020).

[51]  Xiaoyong He, Feng Liu, Fangting Lin, and Wangzhou Shi, "Tunable 3D Dirac-semimetals supported mid-IR hybrid plasmonic waveguides," Opt. Lett. 46, 472-475 (2021).

[52]  Xiaoyong He, Feng Liu, Fangting Lin, and Wangzhou Shi, "Tunable terahertz Dirac semimetal metamaterials," J. Phys. D: Appl. Phys. 54 235103 (2021).

[53]  Jun Peng, Xiaoyong He, Chenyuyi Shi, Jin Leng, Fangting Lin, Feng Liu, Hao Zhang, and Wangzhou Shi, "Investigation of graphene supported terahertz elliptical metamaterials," Physica E, 124, 114309 (2020).

[54]  A. Karvounis, B. Gholipour, K. MacDonald, N. Zheludev, All-dielectric phase-change reconfigurable metasurface, Appl.Phys.Lett. 109(5), 051103, (2016)

[55]  M. Dicken, K. Aydin, I. Pryce, L. Sweatlock, E. Boyd , S. Walavalkar, J. Ma, H. Atwater, Frequency tunable near-infrared metamaterials based on VO2 phase transition, Opt. Express 17(20), 18330 (2009)

[56]  M. Islam, J. Sultana, M. Biabanifard, Z. Vafapour, M. Nine, A. Dinovitser, C. Cordeiro, B. Ng, and D. Abbott, Tunable localized surface Plasmon grapheme metasurface for multiband superabsorption and terahertz sensing, Carbon, 158, 559 (2020)

[57]  Erez Hasman, Vladimir Kleiner, Gabriel Biener, and Avi Niv, Polarization dependent focusing lens by use of quantized

Pancharatnam–Berry phase diffractive optics, Appl.Phys.Lett. 82(3), 328 (2003).

[58]  Si Jia Li, Yun Bo Li, Lei Zhang, Zhang Jie Luo, Bo Wen Han, Rui Qi Li, Xiang Yu Cao, Qiang Cheng, and Tie Jun Cui, Programmable Controls to Scattering Properties of a Radiation Array. Laser & Photonics Reviews, 15, 2000449 (2021).

[59]  Si Jia Li, Yun Bo Li, He Li, Zheng Xing Wang, Chen Zhang, Ze Xu Guo, Rui Qi Li, Xiang Yu Cao, Qiang Cheng, and Tie Jun Cui, A Thin Self-Feeding Janus Metasurface for Manipulating Incident Waves and Emitting Radiation Waves Simultaneously. Ann. Phys. (Berlin) 2020, 2000020.

[60]  Chu, H., Zhang, H., Zhang, Y. et al. Invisible surfaces enabled by the coalescence of anti-reflection and wavefront controllability in ultrathin metasurfaces. Nat. Commun. 12, 4523 (2021).

[61]  Shi-Wang Fan, Sheng-Dong Zhao, Liyun Cao, Yifan Zhu, A-Li Chen, Yan-Feng Wang, Krupali Donda, Yue-Sheng Wang, and Badreddine Assouar, "Reconfigurable curved metasurface for acoustic cloaking and illusion," Phys. Rev. B 101, 024104 (2020).

[62]  Qian, C., Zheng, B., Shen, Y. et al. Deep-learning-enabled self-adaptive microwave cloak without human intervention. Nat. Photonics 14, 383–390 (2020).

[63]  Liyi Hsu, Abdoulaye Ndao, and Boubacar Kanté "Broadband and linear polarization metasurface carpet cloak in the visible," Opt. Lett. 44, 2978-2981 (2019).

[64]  Yijia Huang, Mingbo Pu, Fei Zhang, Jun Luo, Xiong Li, Xiaoliang Ma, and Xiangang Luo," Broadband Functional Metasurfaces: Achieving Nonlinear Phase Generation toward Achromatic Surface Cloaking and Lensing," Advanced Optical Materials, 7, 1801480 (2019).

[65]  M. Zhang, M. Pu, F. Zhang, Y. Guo , Q. He , X. Ma, Y. Huang, X. Li, H. Yu , X. Luo, Plasmonic Metasurfaces or Switchable Photonic Spin-Orbit Interactions Based on Phase Change Materials, Adv. Sci 5(11), 1800835 (2018).